\documentclass[10pt,draft]{dis03}
\usepackage{epsf,amsmath}

\begin{document}

\title{SOME NEWS ON SPIN PHYSICS\footnote{Invited plenary talk
presented at the ``XI$^{\mathrm{th}}$ International Workshop on 
Deep Inelastic Scattering (DIS 2003)'', St. Petersburg, Russia,
April 23-27, 2003.}}

\author{Werner Vogelsang\\
RIKEN-BNL Research Center and BNL Nuclear Theory, \\
Brookhaven National Laboratory,
Upton, NY 11973, U.S.A. \\
E-mail: wvogelsang@bnl.gov }

\maketitle

\vspace*{-5cm}
\begin{flushright}
BNL-NT-03/28 \\
RBRC-338 \\
\end{flushright}
\vspace*{3cm}

\begin{abstract}
\noindent We briefly review  some of the recent developments 
in QCD spin physics.
\end{abstract}

\section{Introduction} 
For many years now, spin physics has played a very prominent
role in QCD. The field has been carried by the hugely successful
experimental program of polarized deeply-inelastic lepton-nucleon 
scattering (DIS), and by a simultaneous tremendous progress in theory. 
This talk summarizes some of the interesting new developments in spin 
physics in the past roughly two years. As we will see, there have yet 
again been exciting new data from polarized lepton-nucleon scattering, 
but also from the world's first polarized $pp$ collider, RHIC. 
There have been very significant advances in theory as well. 
It will not be possible to cover all developments. 
I will select those topics that may be of particular interest
to the attendees of a conference in the ``DIS'' series. 

\section{Nucleon helicity structure}
\subsection{What we have learned so far}
\begin{figure}[!h]
\vspace*{6.cm}
\begin{center}
\includegraphics{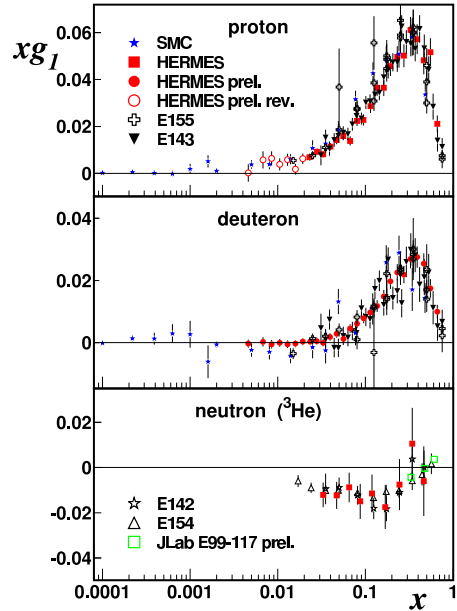}
\vspace*{2.0cm}
\caption[*]{Data on the spin structure function
$g_1$, as compiled and shown in \cite{US02}. \label{fig1}}
\end{center}
\end{figure}

Until a few years ago, polarized inclusive DIS played the dominant 
role in QCD spin physics \cite{hv}. 
At the center of attention was the nucleon's 
spin structure function $g_1(x,Q^2)$. Fig.~\ref{fig1} shows a recent
compilation~\cite{US02} of the world data on $g_1(x,Q^2)$. 
These data have provided much interesting information about the nucleon 
and QCD. For example, they have given direct access to the 
helicity-dependent parton distribution functions of the nucleon,
\begin{equation}
\Delta f(x,Q^2)=f^+ - f^- \label{eq1}\; .
\end{equation}
Polarized DIS actually measures the combinations
$\Delta q+\Delta \bar{q}$. From $x\to 0$ extrapolation of the 
structure functions for proton and neutron targets
it has been possible to test and confirm the Bjorken sum 
rule \cite{bj}. Polarized DIS data, when combined with input from 
hadronic $\beta$ decays, have allowed to extract the -- unexpectedly 
small -- nucleon's axial charge $\sim\,\langle P|\bar{\psi} \,
\gamma^{\mu}\, \gamma^5 \, \psi |P\rangle$, which to lowest order 
unambiguously coincides with the quark spin contribution to the 
nucleon spin \cite{hv}. 

\vspace*{-1.6cm}
\subsection{Things we would like to know}
The results from polarized inclusive DIS have also
led us to identify the next important 
goals in our quest for understanding the spin structure
of the nucleon. The measurement of gluon
polarization $\Delta g=g^+-g^-$ rightly is a 
main emphasis at several experiments in spin physics 
today, since $\Delta g$ could be a major contributor to 
the nucleon spin. Also, more detailed understanding of polarized
quark distributions is clearly needed; for example, 
we would like to know about flavor symmetry breakings in
the polarized nucleon sea, details about strange quark
polarization, and also about the small-$x$ and large-$x$
behavior of the densities. Again, these questions are being
addressed by current experiments. Finally, we would like
to find out how much orbital angular momentum quarks and 
gluons contribute to the nucleon spin. Ji showed \cite{ji} 
that their total angular momenta may be extracted 
from deeply-virtual Compton scattering, which
has sparked much experimental activity also in this area.

\subsection{Current experiments in high-energy spin physics} 
There are several lepton-nucleon scattering fixed-target 
experiments around the world with dedicated spin physics 
programs. This will not be a complete list; 
I will mention only those that play a role in this talk. {\sc Hermes} 
at DESY uses {\sc Hera}'s 27.5~GeV polarized electron beam 
on polarized targets. They have just completed a run with a transversely 
polarized target. Semi-inclusive DIS (SIDIS) measurements are one 
particular strength of {\sc Hermes}. {\sc Compass} at CERN uses 
a 160~GeV polarized muon beam. Their main emphasis is measuring gluon 
polarization; they have completed their first run. There is also 
a very large spin program at Jefferson Lab, involving several 
experiments. Large-$x$ structure functions 
and the DVCS reaction are just two of many objectives 
there. Finally, an experiment E161 at SLAC aims at measuring 
$\Delta g$ in photoproduction, but has unfortunately been put on hold 
awaiting funding. For the more distant future, there are plans to 
develop a polarized electron-proton {\it collider} at BNL, 
eRHIC \cite{eRHIC}.

A new milestone has been reached in spin physics by the advent
of the first polarized proton-proton collider, RHIC at BNL. 
By now, two physics runs with polarized protons colliding 
at $\sqrt{s}=200$~GeV have been completed, and exciting first results 
are emerging. We will see one example toward the end of this
talk. All components crucial for the initial phase of the spin 
program with beam polarization up to 50\% are in place \cite{bland}. 
This is true for the accelerator (polarized source, Siberian snakes, 
polarimetry by proton-Carbon elastic scattering) as well as for 
the detectors. RHIC brings to collision 55 bunches with a polarization 
pattern, for example, $\;\ldots++--++ 
\ldots\;$ in one ring and $\;\ldots+-+-+- \ldots\;$ in the other, 
which amounts to collisions with different spin combinations every 
106~nsec. It has been possible to maintain polarization for about 
10 hours. There is still need for improvements in polarization and 
luminosity for future runs. The two larger RHIC experiments,
{\sc Phenix} and {\sc Star}, have dedicated spin programs focusing
on precise measurements of $\Delta g$, quark polarizations
by flavor, phenomena with transverse spin, and many others.

\subsection{Accessing gluon polarization $\Delta g$}
As mentioned above, the measurement of $\Delta g$ is a 
main goal of several experiments. The gluon density
affects the $Q^2$-evolution of the structure function 
$g_1(x,Q^2)$, but the limited lever arm in $Q^2$ available 
so far has left $\Delta g$ virtually unconstrained.  One way to 
access $\Delta g$ in lepton-nucleon scattering is therefore to 
look at a less inclusive final state that is particularly 
sensitive to gluons in the initial state. One channel, to be 
investigated by {\sc Compass} in particular, is heavy-flavor 
production via the photon-gluon fusion process \cite{compass}.
An alternative reaction is $ep\to h^+ h^- X $, where the
two hadrons in the final state have large transverse
momentum \cite{compass,bravar}. 

\begin{figure}[!h]
\vspace*{3.6cm}
\begin{center}
\includegraphics{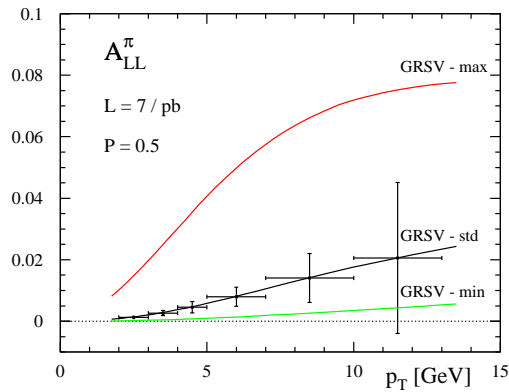}
\vspace*{1.8cm}
\caption[*]{NLO predictions \cite{jssv} for the spin asymmetry
in $pp\to \pi X$ at RHIC, for various $\Delta g$. \label{fig2}}
\end{center}
\end{figure}

RHIC will likely dominate the measurements of $\Delta g$.
Several different processes will be investigated \cite{rhicrev} 
that are sensitive to gluon polarization: high-$p_T$ prompt photons 
$pp\to \gamma  X $, jet or hadron production $pp\to {\rm jet}X$, 
$pp\to h X$, and heavy-flavor production $pp\to (Q\bar{Q}) X$.
In addition, besides the current $\sqrt{s}=200$~GeV, also 
$\sqrt{s}=500$~GeV will be available at a later stage. All
this will allow to determine $\Delta g(x,Q^2)$ in various
regions of $x$, and at different scales. One can compare the
$\Delta g$ extracted in the various channels, and
hence check its universality implied by factorization
theorems. In this way, we will also likely learn a lot
more about high-$p_T$ reactions in QCD. We emphasize that
for all the reactions relevant at RHIC we now know the next-to-leading
order (NLO) QCD corrections to the underlying hard scatterings
of polarized partons \cite{jssv}. This significantly improves
the theoretical framework, since it is known from experience 
with the unpolarized case that the corrections are 
indispensable in order to arrive at quantitative predictions
for hadronic cross sections. For instance, the
dependence on factorization and renormalization scales in the
calculation is much reduced when going to NLO. Therefore, only with 
knowledge of the NLO corrections will one be able to
extract $\Delta g$ reliably. Figure~\ref{fig2} shows NLO 
predictions \cite{jssv} for the double-spin asymmetry $A_{\mathrm{LL}}$ 
for the reaction $pp\to \pi X$ at RHIC, using various different 
currently allowed parameterizations \cite{grsv} of $\Delta g(x,Q^2)$. It
also shows the statistical errors bars expected for
a measurement by {\sc Phenix}\footnote{Very recently, first 
results for $A_{\mathrm{LL}}$ in $pp\to \pi X$ with lower
polarization and luminosity were reported by 
{\sc Phenix} \cite{dubna}.} under the assumption of 50\% beam 
polarizations and 7/pb integrated luminosity. It is evident 
that the prospects for determining $\Delta g$ in this reaction, and
in related ones, are excellent. We stress that 
{\sc Phenix} has recently presented a measurement of the unpolarized
high-$p_T$ $\pi^0$ cross section \cite{phenixpi0} that agrees well 
with an NLO perturbative-QCD calculation over the whole range
of $p_T$ accessed. This provides confidence that the theoretical 
hard scattering framework used for Fig.~\ref{fig2} is indeed 
adequate.

\vspace*{-1.3cm}
\subsection{Further information on quark polarizations}
\begin{figure}[!h]
\vspace*{6.2cm}
\begin{center}
\includegraphics{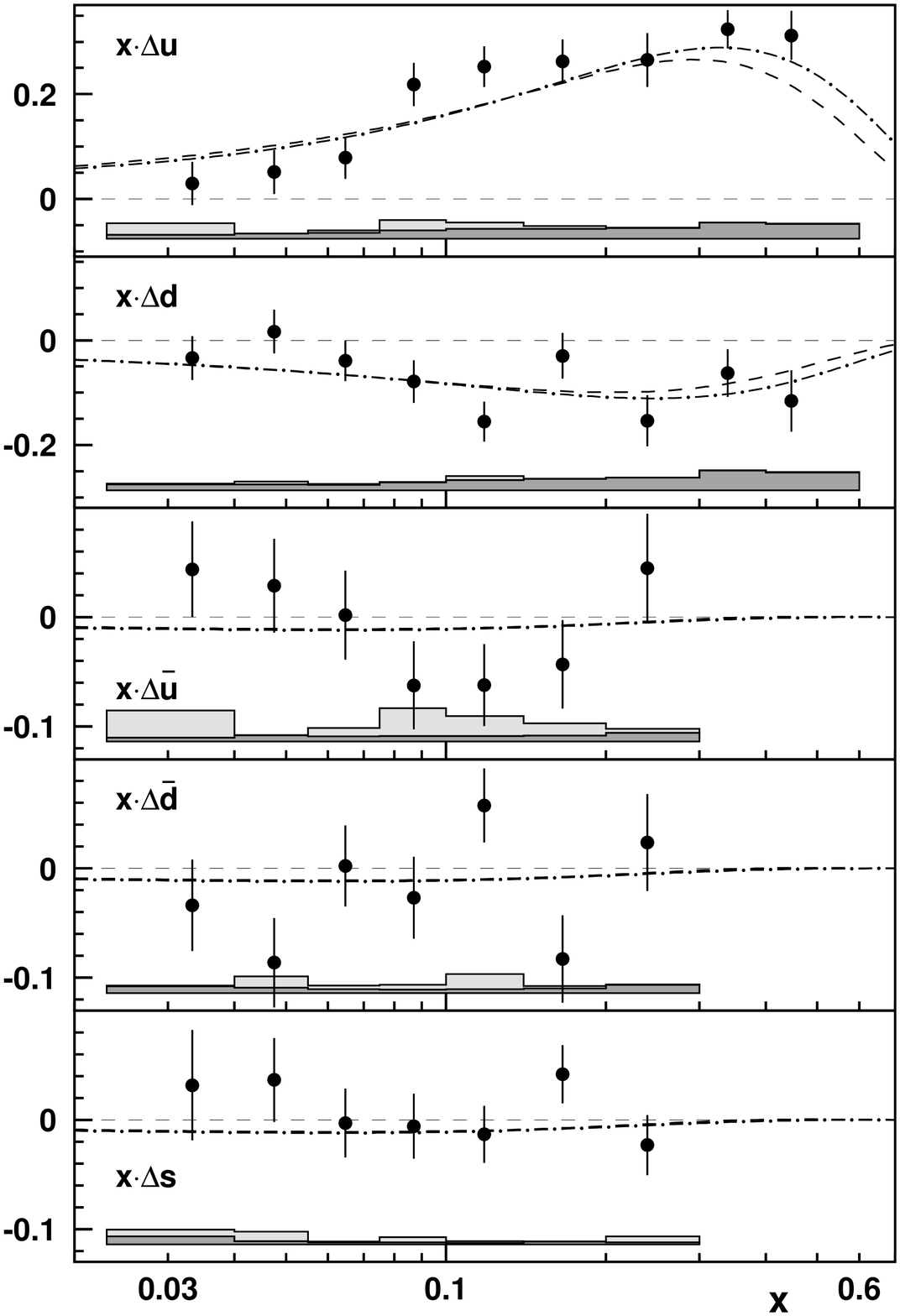}
\vspace*{3.3cm}
\caption[*]{Recent {\sc Hermes} results \cite{hermesdq} 
for the quark and antiquark
polarizations extracted from semi-inclusive DIS.  \label{fig3}}
\end{center}
\end{figure}

As mentioned earlier, inclusive DIS via photon exchange 
only gives access to the combinations $\Delta q+\Delta 
\bar{q}$. There are at least two ways to distinguish between 
quark and antiquark polarizations, and also to achieve a 
flavor separation. Semi-inclusive measurements in DIS 
are one possibility, explored by SMC \cite{smc} and, more recently
and with higher precision, by {\sc Hermes} \cite{hermesdq}. 
One detects a hadron in the final state, so that instead of $\Delta q+
\Delta \bar{q}$ the polarized DIS cross section becomes 
sensitive to $\;\Delta q(x)\,D_q^h(z) + \Delta  \bar{q}(x)\,
D_{\bar{q}}^h (z)\;$, for a given quark flavor. Here, the $D_i^h(z)$ are 
fragmentation functions, with $z=E^h/\nu$. Fig.~\ref{fig3} shows the 
latest results on the flavor separation by {\sc Hermes} 
\cite{hermesdq}, obtained from their LO Monte-Carlo code 
based ``purity'' analysis. Within the still fairly large 
uncertainties, they are not inconsistent with the large negative 
polarization of $\Delta\bar{u}=\Delta\bar{d}=\Delta \bar{s}$ in 
the sea that has been implemented in many determinations of polarized 
parton distributions from inclusive DIS data \cite{grsv,bb} (see curves 
in Fig.~\ref{fig3}). On the other hand, there is no evidence either 
for a large negative strange quark polarization. For the region 
$0.023<x<0.3$, the extracted 
\vspace*{-1.0cm}
$\Delta s$ integrates \cite{hermesdq} to 
the value $+0.03\pm 0.03 \,\mathrm{(stat.)}\,\pm 0.01 \,\mathrm{(sys.)}$,
while analyses of inclusive DIS prefer an integral of about -0.025. 
There is much theory activity currently on SIDIS, focusing also
on possible systematic improvements to the analysis method employed 
in \cite{hermesdq}, among them
NLO corrections, target fragmentation, and higher twist
contributions \cite{sidis}. We note that at RHIC \cite{rhicrev}
one will use $W^{\pm}$ production to determine $\Delta u,\Delta \bar{u},
\Delta d,\Delta \bar{d}$ with good precision, making use of
parity-violation. Comparisons of such data taken at much higher 
scales with those from SIDIS will be extremely interesting. 

\begin{figure}[!h]
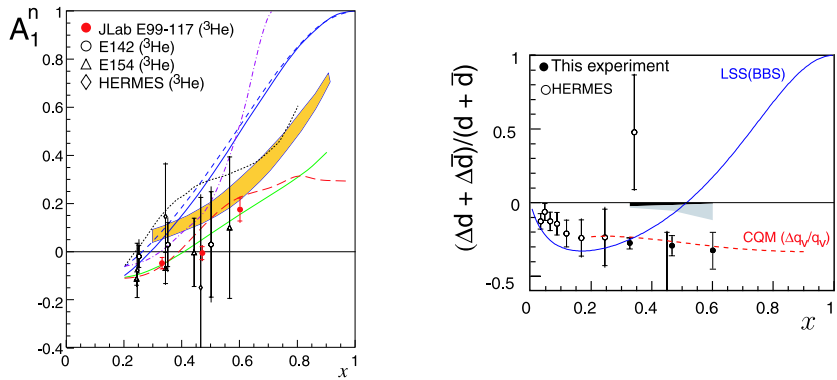

\vspace*{4.9cm}
\begin{center}
\includegraphics{res_a1_zein.epsi}
\includegraphics{deltad_new.epsi}
\caption[*]{Left: Recent data on $A_1^n$ from the E99-117 experiment
\cite{e99117}. Right: extracted polarization asymmetry for $d+\bar{d}$. 
\label{fig4}}
\end{center}
\end{figure}

New interesting information on the polarized quark densities has also recently
been obtained at high $x$. The Hall A collaboration at JLab has 
published their data for the neutron asymmetry $A_1^n$ \cite{e99117}, 
shown in Fig.~\ref{fig4} (left). The new
data points show a clear trend for $A_1^n$ to turn positive
at large $x$. Such data are valuable because the
valence region is a particularly useful testing ground for
models of nucleon structure. The right panel of Fig.~\ref{fig4}
shows the extracted polarization asymmetry for $d+\bar{d}$. The data
are consistent with constituent quark models \cite{cqm}
predicting $\Delta d/d\to -1/3$ at large $x$, while 
``hadron helicity conservation'' predictions based on perturbative 
QCD and the neglect of quark orbital angular
momentum \cite{hhc} give $\Delta d/d\to 1$ and tend to deviate from 
the data, unless the convergence to 1 sets in very late.

\vspace*{-1.2cm}
\section{Transverse-spin phenomena}
\vspace*{-0.2cm}
\subsection{Transversity} 
\begin{figure}[!h]
\vspace*{7.cm}
\begin{center}
\includegraphics{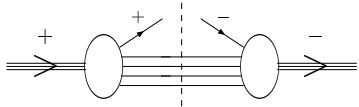}
\vspace*{-5.4cm}
\caption[*]{Transversity in helicity basis. \label{fig5}}
\vspace*{-8mm}
\end{center}
\end{figure}

Besides the unpolarized and the helicity-dependent
densities, there is a third set of twist-2 parton 
distributions, transversity \cite{rs}. In analogy with Eq.~(\ref{eq1})
they measure the net number (parallel minus antiparallel) 
of partons with transverse polarization in a transversely
polarized nucleon:
\begin{equation}
\delta f(x,Q^2)=f^{\uparrow} - f^{\downarrow} \label{eq2}\; .
\end{equation}
In a helicity basis, one finds \cite{rs} that transversity 
corresponds to a helicity-flip structure, as shown in Fig.~\ref{fig5}.
This precludes a gluon transversity distribution at 
leading twist. It also makes transversity
a probe of chiral symmetry breaking in QCD \cite{collins}: 
perturbative-QCD interactions preserve chirality,
and so the helicity flip required to make transversity non-zero must
primarily come from soft non-perturbative interactions for which 
chiral symmetry is broken. 

\begin{figure}[!h]
\vspace*{-0.7cm}
\begin{center}
\includegraphics{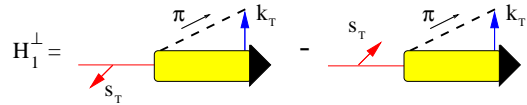}
\vspace*{2.cm}
\caption[*]{The Collins function. \label{fig6}}
\vspace*{-0.7cm}
\end{center}
\end{figure}

Measurements of transversity are not straightforward. Again
the fact that perturbative interactions in the Standard
Model do not change chirality (or, for massless quarks, helicity) 
means that inclusive DIS is not useful. Collins, however, 
showed \cite{coll93} that properties of fragmentation 
might be exploited to obtain a ``transversity polarimeter'': a pion produced 
in fragmentation will have some transverse momentum with respect to the
fragmenting parent quark. There may then be a correlation
of the form $\;i\vec{S}_T \cdot  (\vec{P}_{\pi} \times \vec{k}_{\perp})$. 
The fragmentation function associated with this correlation
is the Collins function. The phase is required by time-reversal 
invariance. The situation is depicted in Fig.~\ref{fig6}. 
The 
Collins function would make a {\it leading-power} \cite{coll93}
contribution to the single-spin asymmetry $A_{\perp}$ in the 
reaction $\;ep^{\uparrow}
\to e\pi X$:
\begin{equation} 
A_{\perp}\propto|\vec{S}_T|
\sin(\phi+\phi_S)\sum_q\,e_q^2
\delta q(x)H_1^{\perp,q}(z) \; ,\label{eq3}
\end{equation}
where $\phi$ ($\phi_S$) is the angle between the lepton plane
and the $(\gamma^* \pi)$ plane (and the transverse target spin).
As is evident from Eq.~(\ref{eq3}), this asymmetry would
allow access to transversity if the Collins functions are
non-vanishing. A few years ago, {\sc Hermes} measured the asymmetry 
for a longitudinally polarized target \cite{hermessp}. 
For finite $Q$, the target spin then has a transverse component
$\propto M/Q$ relative to the direction of the virtual photon, 
and the effect may still be there, even though it is now only one
of several ``higher twist'' contributions \cite{abko}. 

\subsection{News on the Sivers function} 
\begin{figure}[!h]
\vspace*{2.8cm}
\begin{center}
\includegraphics{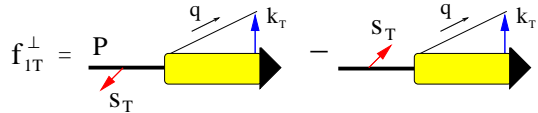}
\vspace*{-1.2cm}
\caption[*]{The Sivers function. \label{fig7}}
\vspace*{-0.8cm}
\end{center}
\end{figure}

If ``intrinsic'' transverse momentum in the fragmentation
process plays a crucial role in the asymmetry for 
$\;ep^{\uparrow} \to e\pi X$, 
a natural question is whether $k_{\perp}$ in the initial 
state can be relevant as well. Sivers suggested \cite{sivers}
that the $k_{\perp}$ distribution of a quark in a transversely
polarized hadron could have an azimuthal asymmetry, 
$\,\vec{S}_T \cdot  (\vec{P} \times \vec{k}_{\perp})$,
as shown in Fig.~\ref{fig7}. 
There is a qualitative difference between the Collins and 
Sivers functions, however. While phases will always arise in
strong interaction final-state fragmentation, one does not 
expect them from initial (stable) hadrons, and the Sivers
function appears to be ruled out by time-reversal invariance
of QCD \cite{coll93}. Until recently,
it was therefore widely believed that origins of single-spin 
asymmetries as in $\;ep^{\uparrow}\to e\pi X$ and other
reactions were more likely to be found in final-state 
fragmentation effects than in initial state parton distributions. 
However, then came a model calculation \cite{bbs} that
found a leading-power asymmetry in $\,ep^{\uparrow}\to e\pi X$
not associated with the Collins effect. It was subsequently 
realized \cite{coll02,bjy,bmp} that the calculation of \cite{bbs}
could be regarded as a model for the Sivers effect. 
It turned out that the original time-reversal argument against 
the Sivers function is invalidated by the presence of the
Wilson lines in the operators defining the parton density.
These are required by gauge invariance and had been neglected
in \cite{coll93}. Under time reversal, however, future-pointing 
Wilson lines turn into past-pointing ones, which changes the 
time reversal properties of the Sivers function and allows
it to be non-vanishing. Now, for a ``standard'', $k_{\perp}$-integrated, 
parton density the gauge link contribution is unity in the $A^+=0$ 
gauge, so one may wonder how it can be relevant
for the Sivers function. The point, however, is that 
for the case of $k_{\perp}$-dependent parton densities, a gauge link
survives even in the light-cone gauge, in a transverse direction
at light-cone component $\xi^-=\infty$ \cite{bjy,bmp}. Thus, time 
reversal indeed does not imply that the Sivers function vanishes. 
The same is true for a function describing transversity in an 
unpolarized hadron \cite{boer}. It is intriguing that these new 
results are based entirely on the Wilson lines in QCD.  

\subsection{Implications for phenomenology}
If the Sivers function is non-vanishing, it will for example 
make a leading-power contribution to $\;ep^{\uparrow}\to e\pi X$,
of the form
\begin{equation} 
A_{\perp}\propto|\vec{S}_T|
\sin(\phi-\phi_S)\;\sum_q\,e_q^2\;
f_{1T}^{\perp ,q}(x)\;D_q^{\pi}(z)  \; .\label{eq4}
\end{equation}
This is in competition with the Collins function contribution,
Eq.~(\ref{eq3}); however, the azimuthal angular dependence
is discernibly different. {\sc Hermes} has just completed a run
with transverse polarization, and preliminary results are 
expected soon. We note that the Collins function may also
be determined separately from an azimuthal asymmetry in
$e^+e^-$ annihilation \cite{dbee}. It was pointed out 
\cite{coll02,bjy,bmp} that comparisons of DIS and the
Drell-Yan process will be particularly interesting:
from the properties of the Wilson lines it follows
that the Sivers functions relevant in DIS and in the 
Drell-Yan process have opposite sign, violating universality
of the distribution functions. This is a striking prediction
awaiting experimental testing. For work on the process 
(in)dependence of the Collins function, see \cite{bmp,metz};
recent model calculations of the function in the context
of the gauge links may be found in \cite{models}.

\begin{figure}[!h]
\vspace*{6.1cm}
\begin{center}
\includegraphics{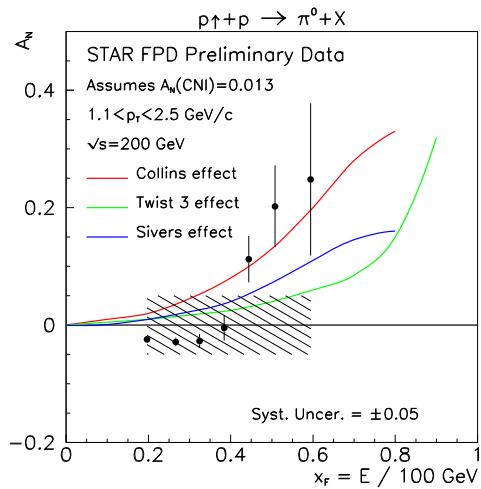}
\vspace*{2mm}
\caption[*]{Recent preliminary results from STAR for
the asymmetry $A_{\mathrm{N}}$ in $pp\to \pi^0X$ in the forward
region \cite{bland}. \label{fig9}}
\vspace*{-1.2cm}
\end{center}
\end{figure}

Originally, the Sivers function was proposed \cite{sivers} as a means
to understand and describe the significant single-spin 
asymmetries $A_{\mathrm{N}}$ observed \cite{anold} in $p^{\uparrow}p
\to \pi X$, with the pion at high $p_T$. These are inclusive 
``left-right'' asymmetries and may be generated by the Sivers 
function from the effects of the quark intrinsic transverse momentum 
$k_{\perp}$ on the partonic hard-scattering which
has a steep $p_T$ dependence. The resulting asymmetry $A_{\mathrm{N}}$ is then 
power-suppressed as $\sim\langle k_{\perp} \rangle /p_T$ in QCD, 
where $\langle k_{\perp} \rangle$ is an average intrinsic transverse
momentum. Similar effects may arise also from the Collins
function. Fits to the available $A_{\mathrm{N}}$ data have been  
performed recently \cite{dalesio}, assuming variously dominance of the Collins 
or the Sivers mechanisms. An exciting new development in the
field is that the {\sc Star} collaboration has presented the 
first data on $p^{\uparrow}p\to \pi X$ from RHIC \cite{bland}. 
The results are shown in Fig.~\ref{fig9}. As one can see,
a large $A_{\mathrm{N}}$ persists to these much higher energies.
Fig.~\ref{fig9} also shows predictions based on 
the Collins and the Sivers effects \cite{dalesio}, 
and on a formalism \cite{qs,koike} that
systematically treats the power-suppression of $A_{\mathrm{N}}$ in terms
of higher-twist parton correlation functions (for a connection
of the latter with the Sivers effect, see \cite{bmp}). 
The STAR data clearly give valuable information already now. For
the future, it will be important to extend the 
measurements to higher $p_T$ where the perturbative-QCD
framework underlying all calculations will become more 
reliable. 

\subsection{Two other developments}
It was recognized some time ago that certain Fourier transforms of 
generalized parton densities with respect to momentum transfer 
give information on the position space distributions of partons 
in the nucleon \cite{ft}. For a transversely polarized nucleon, 
one then expects \cite{mb} a distortion of the parton distributions in
the transverse plane, which could provide an intuitive physical picture 
for the origins of single-spin asymmetries. 

We finally note that {\it double}-transverse spin asymmetries 
$A_{\mathrm{TT}}$ in $pp$ scattering offer another possibility to access
transversity. Candidate processes are Drell-Yan, prompt photon, 
and jet production. Recently, the NLO corrections
to $p^{\uparrow}p^{\uparrow}\to \gamma X$ have been 
calculated \cite{msv}. The results show that $A_{\mathrm{TT}}$
is expected rather small at RHIC. 

\section*{Acknowledgements}
I am grateful to the organizers of DIS~2003 for their 
invitation. I thank D.~Boer, G.\ Bunce, M.\ Grosse-Perdekamp, 
S.\ Kretzer, Z.\ Meziani, G.\ Rakness, M.\ Stratmann
for very useful discussions and help, and RIKEN, Brookhaven National 
Laboratory and the U.S. Department of Energy (contract number 
DE-AC02-98CH10886) for providing the facilities essential for 
the completion of this work.

\end{document}